\documentclass{aa}
\usepackage{graphics,epsfig,lscape,times}

\usepackage{natbib}
\bibpunct{(}{)}{;}{a}{}{,}

\newcommand{\oergcm}[1]{$10^{#1}$ erg cm$^{-2}$ s$^{-1}$}

\newcommand{\hcm}[1]{$\cdot 10^{#1}$ cm$^{-2}$}

\newcommand{\expo}[1]{$\cdot 10^{#1}$}
\newcommand{\oexpo}[1]{$10^{#1}$}
\newcommand{\nh}{N$_{\rm H}$}

\newcommand{\eline}{\hbox{E$_{\rm line}$}}

\newcommand{\eqw}{\hbox{EW}}

\newcommand{\pdot}{$\dot{\rm P}$}
\newcommand{\xmm}{\hbox{XMM-Newton}}
\newcommand{\rxa}{\hbox{\object{RX\,J0420.0$-$5022}}}
\newcommand{\rxb}{\hbox{\object{RX\,J0720.4$-$3125}}}
\newcommand{\rxc}{\hbox{\object{RX\,J0806.4$-$4123}}}
\newcommand{\rxd}{\hbox{\object{1RXS\,J130848.6+212708}}}
\newcommand{\rxe}{\hbox{\object{RX\,J1605.3+3249}}}
\newcommand{\rxf}{\hbox{\object{RX\,J1856.4$-$3754}}}
\newcommand{\rbs}{\hbox{\object{RBS1223}}}

\begin{document}

\title{The isolated neutron star X-ray pulsars \rxa\ and \rxc: new X-ray and optical observations
       \thanks{Based on observations with XMM-Newton, an ESA Science Mission
               with instruments and contributions directly funded by ESA Member
               states and the USA (NASA). Optical observations were performed at the European
           Southern Observatory, ESO programme 66.D-0128(A).}
}

\author{F.~Haberl\inst{1} \and C.~Motch\inst{2} \and V.E.~Zavlin\inst{1,2}
                          \and K.~Reinsch\inst{3} \and B.T.~G\"ansicke\inst{4}
                          \and M.~Cropper \inst{5}
                          \and A.D.~Schwope\inst{6} \and R.~Turolla\inst{7} \and S.~Zane\inst{5}}

\titlerunning{The isolated neutron star X-ray pulsars \rxa\ and \rxc}
\authorrunning{Haberl et al.}

\offprints{F. Haberl, \email{fwh@mpe.mpg.de}}

\institute{Max-Planck-Institut f\"ur extraterrestrische Physik,
           Giessenbachstra{\ss}e, 85748 Garching, Germany
          \and
          Observatoire Astronomique, CNRS UMR 7550, 11 Rue de l'Universit\'e,
          67000 Strasbourg, France
          \and
          Universit\"ats-Sternwarte, Geismarlandstr. 11, 37083 G\"ottingen, Germany
          \and
	  Department of Physics, University of Warwick, Coventry CV4 7AL, UK
	  \and
          Mullard Space Science Laboratory, University College London,
          Holmbury St. Mary, Dorking, Surrey RH5 6NT, UK
          \and
          Astrophysikalisches Institut Potsdam, An der Sternwarte 16,
          14482 Potsdam, Germany
          \and
          Dipartimento di Fisica, Universit\'a di Padova,
           Via Marzolo 8, 35131 Padova, Italy}

\date{Received 16 March 2004 / Accepted 18 May 2004}

\abstract{ We report on the analysis of new X-ray data obtained
with XMM-Newton and Chandra from two ROSAT-discovered X-ray dim
isolated neutron stars (XDINs). \rxc\ was observed with XMM-Newton
in April 2003, 2.5 years after the first observation. The EPIC-pn
data confirm that this object is an X-ray pulsar with 11.371 s
neutron star spin period. The X-ray spectrum is consistent with
absorbed black-body emission with a temperature kT = 96 eV and
\nh\ = 4\hcm{19} without significant changes between the two
observations. Four XMM-Newton observations of \rxa\ between
December 2002 and July 2003 did not confirm the 22.7 s pulsations
originally indicated in ROSAT data, but clearly reveal a 3.453 s
period. A fit to the X-ray spectrum using an absorbed black-body
model yields kT = 45 eV, the lowest value found from the small
group of XDINs and \nh\ = 1.0\hcm{20}. Including a broad
absorption line improves the quality of the spectral fits considerably 
for both objects and may indicate the presence of absorption
features similar to those reported from \rbs, \rxe\ and \rxb.
For both targets we derive accurate X-ray positions from the
Chandra data and present an optical counterpart candidate for
\rxa\ with B = 26.6$\pm$0.3 mag from VLT imaging.
}

\maketitle

\section{Introduction}

The two soft X-ray sources \rxc\ and \rxa\ belong to the group of X-ray dim isolated neutron
stars (XDINs) discovered in the ROSAT all-sky survey data
\citep[for recent reviews see][]{2000PASP..112..297T,2001xase.conf..244M,haberl2004COSPAR}.
\rxc\ was discovered by \citet{1998AN....319...97H} in a dedicated search for XDINs applying
spectral hardness ratio selection criteria to the ROSAT survey source catalogue restricted to the
galactic plane. Two short ROSAT pointed PSPC observations covered the source serendipitously at an
off-axis angle of 31\arcmin\ and the obtained spectra were consistent with a black-body model with
temperature kT = 78$\pm$7 eV. No optical counterpart was identified with a limiting B
magnitude of $\sim$24. A first XMM-Newton observation as part of the telescope scientist
guaranteed time revealed a possible candidate for the neutron star spin period of 11.37 s,
formally detected at a 3.5$\sigma$ level in the EPIC-pn data
\citep[][hereafter HZ02]{2002A&A...391..571H}.

The faintest object in X-rays among the XDINs is \rxa, which
was selected initially due to a mis-identification with a
nearby galaxy. Follow-up ROSAT PSPC and HRI observations and ESO-NTT images, which revealed
no optical counterpart brighter than B = 25.25, ruled out possible kinds of known X-ray emitters
other than an isolated neutron star \citep{1999A&A...351L..53H}. The X-ray spectrum -
consistent with a Planckian shape as observed from the other XDINs - obtained from the
short PSPC pointed observation allowed only coarse estimates of black-body temperature and
interstellar absorption, but indicated that it is one of the XDINs with lowest temperature. The PSPC data
also showed evidence for periodic modulation of the X-ray flux with a period
of 22.7 s suggesting that \rxa\ is a X-ray pulsar similar to
\rxb\ \citep{1997A&A...326..662H}, the only
pulsar known among the XDINs at that time.

\begin{table*}
\caption[]{Chandra ACIS and XMM-Newton EPIC observations.}
\begin{center}
\begin{tabular}{lllcccccc}
\hline\hline\noalign{\smallskip}
\multicolumn{1}{l}{Target} &
\multicolumn{1}{l}{Instrument} &
\multicolumn{1}{c}{Read-out} &
\multicolumn{1}{c}{Filter} &
\multicolumn{1}{c}{Sat. Revol./} &
\multicolumn{3}{c}{Observation} &
\multicolumn{1}{c}{Exp.} \\
 &
 &
\multicolumn{1}{c}{Mode} &
 &
\multicolumn{1}{c}{Obs.-ID} &
\multicolumn{1}{c}{Date} &
\multicolumn{1}{c}{Start} &
\multicolumn{1}{c}{End (UT)} &
\multicolumn{1}{c}{[ks]} \\

\noalign{\smallskip}\hline\noalign{\smallskip}
\rxc & EPIC-MOS1/2  & FF, 2.6 s & Thin & 168 & Nov.  8, 2000 & 13:43 & 18:47 & 18.0 \\
     & EPIC-pn      & FF, 73 ms & Thin &     &               & 14:24 & 18:53 & 15.6 \\
     & RGS1/2       & Spectro   & --   &     &               & 13:35 & 18:54 & 19.1 \\
\noalign{\smallskip}
     & ACIS-I       & FF, 3.2 s & $-$  & 500239 & Feb. 21, 2002 & 03:33 & 08:55 & 17.9 \\
\noalign{\smallskip}
     & EPIC-MOS1/2  & FF, 2.6 s & Thin & 618 & Apr. 24, 2003 & 14:18 & 21:10 & 24.6 \\
     & EPIC-pn      & FF, 73 ms & Thin &     &               & 14:44 & 21:10 & 22.7 \\
     & RGS1/2       & Spectro   & --   &     &               & 13:49 & 21:11 & 26.4 \\
\noalign{\smallskip}\hline\noalign{\smallskip}
\rxa & ACIS-S       & FF, 3.2 s & $-$  & 500238 & Nov. 11, 2002 & 21:58 & 03:47 & 19.6 \\
\noalign{\smallskip}
     & EPIC-MOS1/2  & FF, 2.6 s & Thin & 560 & Dec. 30, 2002 & 03:39 & 09:43 & 21.7 \\
     & EPIC-pn      & FF, 73 ms & Thin &     &               & 04:01 & 09:43 & 20.0 \\
\noalign{\smallskip}
     & EPIC-MOS1/2  & FF, 2.6 s & Thin & 561 & Dec. 31, 2002 & 21:55 & 03:59 & 21.7 \\
     & EPIC-pn      & FF, 73 ms & Thin &     &               & 22:17 & 03:59 & 20.0 \\
\noalign{\smallskip}
     & EPIC-MOS1/2  & FF, 2.6 s & Thin & 570 & Jan. 19, 2003 & 16:42 & 22:55 & 22.2 \\
     & EPIC-pn      & FF, 73 ms & Thin &     &               & 17:05 & 22:55 & 20.5 \\
\noalign{\smallskip}
     & EPIC-MOS1/2  & FF, 2.6 s & Thin & 664 & Jul. 25, 2003 & 21:22 & 03:26 & 21.7 \\
     & EPIC-pn      & FF, 73 ms & Thin &     &               & 21:45 & 03:26 & 20.0 \\
\noalign{\smallskip}\hline
\end{tabular}
\end{center}
\label{xray-obs}
\end{table*}

Here we present the results from the analysis of new XMM-Newton and Chandra data from \rxc\ and
\rxa. We derive accurate X-ray positions utilizing the imaging capabilities of Chandra and
derive X-ray spectral parameters from the high statistical quality spectra collected by the
EPIC-pn instrument on board XMM-Newton. We discuss the measured properties in comparison with
the other known XDINs.

\section{Chandra and XMM-Newton observations}

Using XMM-Newton \citep{2001A&A...365L...1J} we observed \rxc\ and \rxa\ during the AO2
guest observer programme.
Here we report on the analysis of the data collected with the European Photon Imaging Camera
based on a pn \citep[EPIC-pn,][]{2001A&A...365L..18S} CCD detector which is mounted behind
one of the three X-ray telescopes \citep{2000SPIE.4012..731A}. We include the first
\xmm\ observation of \rxc\ (HZ02) from the guaranteed time programme for a
comparative study. 
Spectra obtained from \rxc\ (\rxa\ is not sufficiently bright) by the Reflection 
Grating Spectrometers \citep[RGS, ][]{2001A&A...365L...7D} are also investigated.
Details of the XMM-Newton observations are summarized in Table~\ref{xray-obs}. 
We do not use the data from the EPIC-MOS \citep[EPIC-MOS1 and -MOS2,][]{2001A&A...365L..27T} CCDs
because of currently un-calibrated changes in the low-energy spectral response.
With Chandra we observed the two neutron stars in Cycle 3 of the guest observer
programme using the ACIS CCD arrays (Table~\ref{xray-obs}). We used the ACIS data to derive
precise X-ray positions. Due to the low statistics (see below) we did not accumulate
energy spectra.

\subsection{X-ray positions}

The best X-ray positions were derived from the Chandra ACIS pre-processed level 2 event
files using the `celldetect' task of the Chandra analysis package CIAO 3.0. To optimize
the signal to noise ratio we accumulated images in the energy bands 0.3-1.0 keV for \rxc\ 
and 0.3-0.7 keV for \rxa\ (see below). \rxc\ was
detected with 986 net source counts (0.3-1.0 keV, corrected for point spread function losses) and we obtained
RA = 08$^{\rm h}$06$^{\rm m}$23\fs40 and Dec = --41\degr22\arcmin30\farcs9 (J2000.0).
For \rxa\ 270 net source counts (0.3-0.7 keV) were found yielding a position
RA = 04$^{\rm h}$20$^{\rm m}$01\fs95 and Dec = --50\degr22\arcmin48\farcs1 (J2000.0).
The 90\% confidence systematic position uncertainty is 0.6\arcsec\ \citep{2000SPIE.4012..650A}.
We compared the X-ray positions of four \rxa\ field sources (detected at relatively large off-axis angles
of 4\arcmin\ - 5.5\arcmin) with counterparts from the USNO-B catalogue and found agreement to
better than 0.5\arcsec\ in each coordinate. We did not find any systematic shift which would allow to 
reduce the systematic position error.
The Chandra position of \rxc\ is consistent with that derived from the XMM-Newton
observation in Nov. 2000 (HZ02).

\subsection{Average EPIC-pn X-ray spectra}

To obtain energy spectra with highest statistical quality we used
the EPIC-pn data and processed the photon event files using a
the XMM-Newton analysis package SAS version
6.0. Single-pixel events from circular regions around the sources
with 30\arcsec\ (\rxa) and 35\arcsec\ (\rxc) radius were extracted
and background was taken from similar nearby source-free areas.
The SAS task `arfgen' was used to calculate effective areas and
ready-made redistribution matrices for EPIC-pn full-frame (FF)
mode were used for the spectral analysis.

\begin{figure*}
\resizebox{8.7cm}{!}{\includegraphics[angle=-90,clip=]{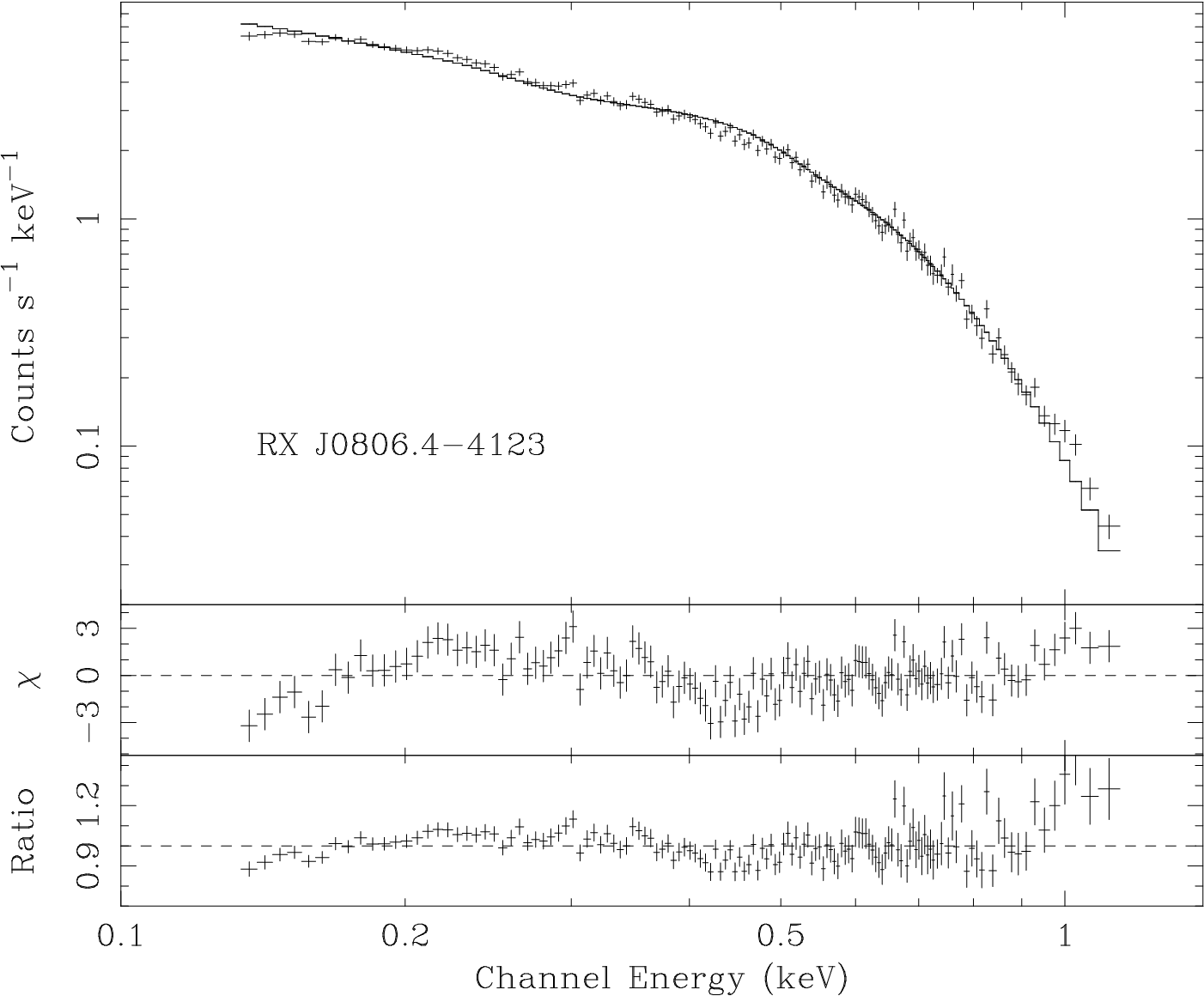}}
\hspace{2mm}
\resizebox{8.7cm}{!}{\includegraphics[angle=-90,clip=]{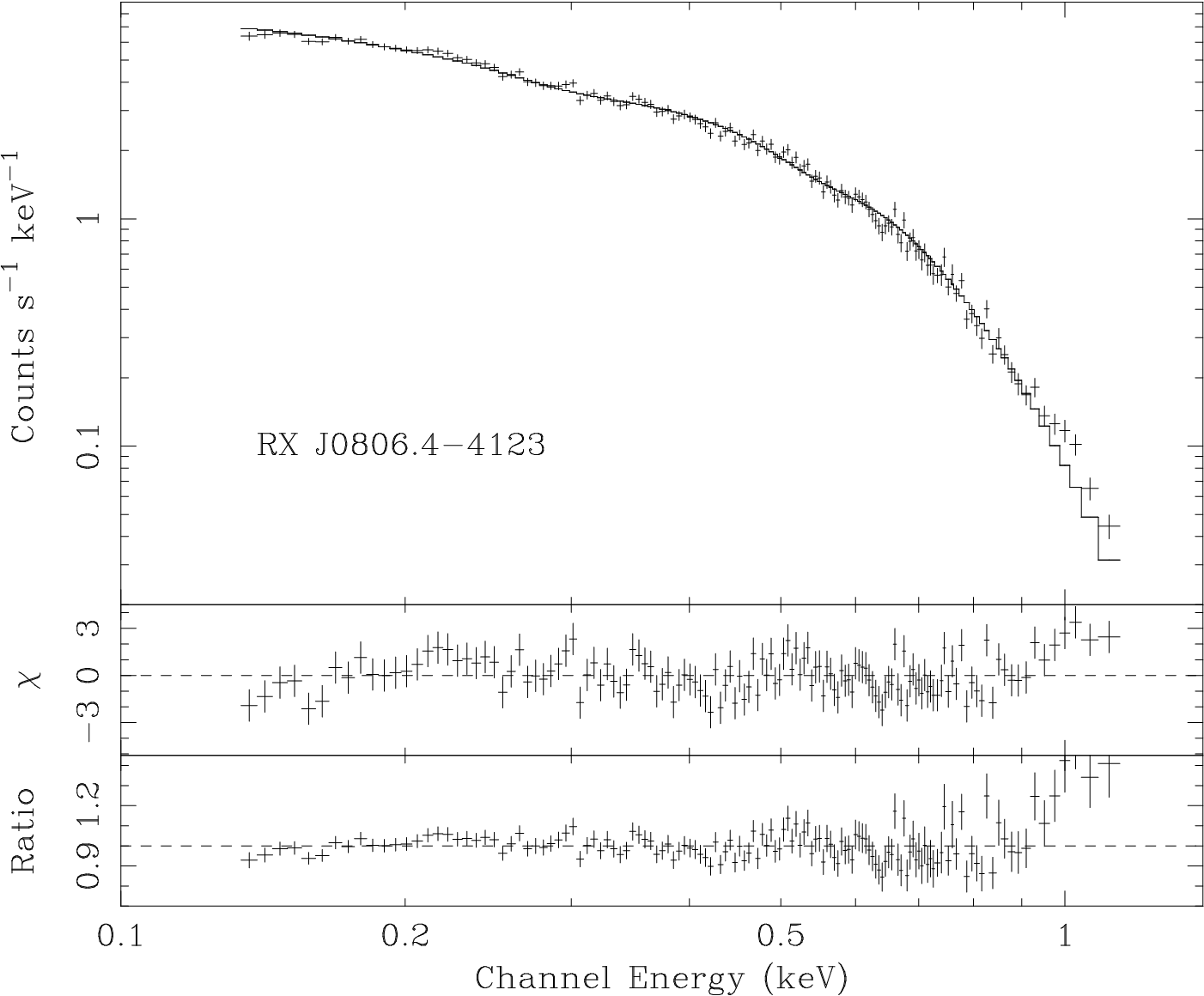}}

\vspace{3mm}
\resizebox{8.7cm}{!}{\includegraphics[angle=-90,clip=]{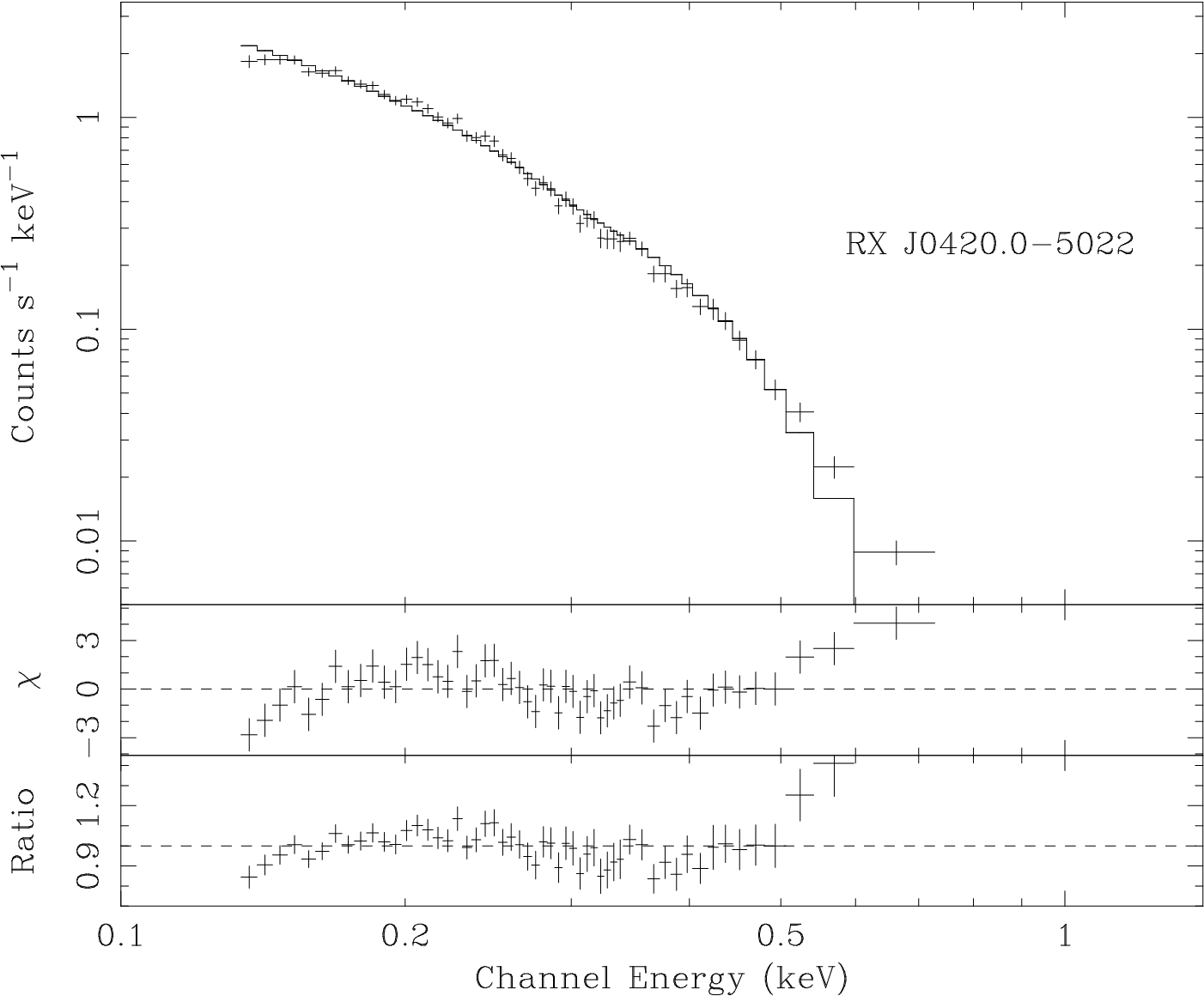}}
\hspace{2mm}
\resizebox{8.7cm}{!}{\includegraphics[angle=-90,clip=]{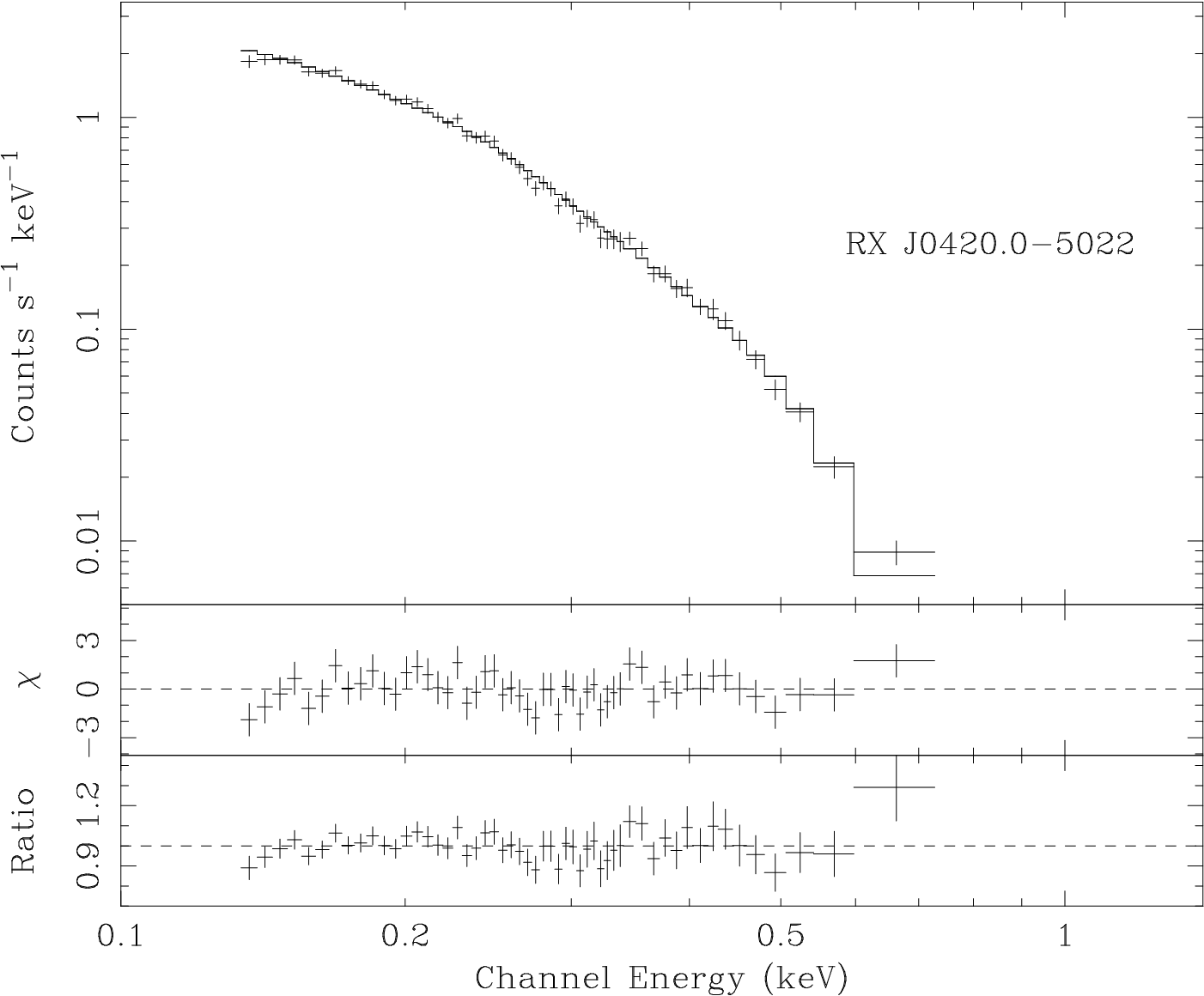}}
\caption{Merged EPIC-pn spectra of \rxc\ and \rxa\ accumulated from all available observations.
         Left: The best fit black-body model with photo-electric absorption is shown as histogram.
     Right: Model fit including an absorption line with Gaussian shape.
         The lower two panels in each case show the fit residuals on the same scale for each target
     (plotted in units of $\sigma$ and as ratio data/model).}
\label{epic-spectra}
\end{figure*}

\begin{table*}
\caption[]{Black-body model fits to the EPIC-pn spectra obtained from individual observations.}
\begin{center}
\begin{tabular}{lcccccc}
\hline\hline\noalign{\smallskip}
\multicolumn{1}{l}{Object} &
\multicolumn{1}{l}{Obs.} &
\multicolumn{1}{c}{Count Rate} &
\multicolumn{1}{c}{kT} &
\multicolumn{1}{c}{\nh} &
\multicolumn{1}{c}{Flux 0.1-2.4 keV} &
\multicolumn{1}{c}{$\chi^2_{\rm r}$/dof} \\

\multicolumn{1}{l}{} &
\multicolumn{1}{l}{} &
\multicolumn{1}{c}{[cts s$^{-1}$]} &
\multicolumn{1}{c}{[eV]} &
\multicolumn{1}{c}{[\oexpo{20}cm$^{-2}$]} &
\multicolumn{1}{c}{\oergcm{-13}} &
\multicolumn{1}{c}{} \\

\noalign{\smallskip}\hline\noalign{\smallskip}
\rxc & 168-pn & 1.780$\pm$0.012 & 95.6$\pm$1.1 & 0.40$\pm$0.11 & 28.8$\pm$0.2 &      \\
     & 618-pn & 1.769$\pm$0.010 & 95.5$\pm$0.9 & =         & 28.7$\pm$0.2 & 1.50/273 \\
\hline\noalign{\smallskip}
\rxa & 560-pn & 0.212$\pm$0.004 & 44.8$\pm$1.8 & 1.01$\pm$0.37 &  4.8$\pm$0.3 &      \\
     & 561-pn & 0.222$\pm$0.004 & =        & =         &  5.1$\pm$0.3 &      \\
     & 570-pn & 0.217$\pm$0.004 & =        & =         &  5.1$\pm$0.3 &      \\
     & 664-pn & 0.212$\pm$0.004 & =        & =         &  4.9$\pm$0.3 & 1.25/157  \\
\noalign{\smallskip}\hline
\end{tabular}
\end{center}

Count rates with 68\% errors correspond to the 0.13-1.2 keV (\rxc) and 0.13-0.9 keV (\rxa)
bands as used for the spectral analysis. Fluxes are corrected for point spread
function losses. Temperatures and fluxes are not corrected for relativistic effects and are
given as measured by a distant observer. Errors on the spectral model parameters
are derived for a 90\% confidence level. \\
``=" denotes fit parameter is a single free parameter common to all spectra of one object.
\label{tab-sresults}
\end{table*}

\subsubsection{Black-body model}

The EPIC-pn spectra obtained from the two XMM-Newton observations of \rxc\ were fit
simultaneously first with an absorbed black-body model.
The column density was treated as a free parameter common to both spectra. Temperature
and normalization were allowed to vary freely for both spectra independently to
allow for possible small gain variations
\citep[see also ][]{2004A&A...419.1077H} and to investigate long-term flux variations. Both
spectra are fully compatible in shape and absolute flux and the derived spectral
parameters, which are summarized in Table~\ref{tab-sresults}, agree within the statistical errors
(90\% confidence for one parameter of interest). In comparison to the fit presented by
HZ02 the up-dated calibration yields reduced residuals and a more
reliable value for the absorption (the \nh\ value derived with the same calibration from the
EPIC-pn spectrum of \rxf\ agrees with that obtained from the Chandra LETGS spectrum).
Because no significant differences are seen in the two spectra, we merged the data and accumulated
a combined spectrum which is presented in Fig.~\ref{epic-spectra}. Again it was modeled
using the absorbed black-body spectrum and the best fit parameters are listed in
Table~\ref{tab-mresults}.

\begin{table*}
\caption[]{Fits to the merged EPIC-pn and RGS spectra.}
\begin{center}
\begin{tabular}{llccccccc}
\hline\hline\noalign{\smallskip}
\multicolumn{1}{l}{Object} &
\multicolumn{1}{l}{Instr.} &
\multicolumn{1}{c}{Phases} &
\multicolumn{1}{c}{Model} &
\multicolumn{1}{c}{kT} &
\multicolumn{1}{c}{\nh} &
\multicolumn{1}{c}{\eline} &
\multicolumn{1}{c}{\eqw} &
\multicolumn{1}{c}{$\chi^2_{\rm r}$/dof} \\

\multicolumn{1}{l}{} &
\multicolumn{1}{l}{} &
\multicolumn{1}{c}{} &
\multicolumn{1}{c}{} &
\multicolumn{1}{c}{[eV]} &
\multicolumn{1}{c}{[\oexpo{20}cm$^{-2}$]} &
\multicolumn{1}{c}{[eV]} &
\multicolumn{1}{c}{[eV]} &
\multicolumn{1}{c}{} \\

\noalign{\smallskip}\hline\noalign{\smallskip}
\rxc & EPIC-pn & all  & black-body     & 95.6$\pm$0.9 & 0.41$\pm$0.09 &    --      &   --       & 1.97/137   \\
     &         & all  & bb + gauss$^1$ & 91.7$\pm$1.4 & 1.12$\pm$0.22 & 460$\pm$15 & $-33\pm$6  & 1.51/135   \\
     & RGS1/2  & all  & black-body     & 104$\pm$4    & 0.41$^2$      &    --	  &   --       & 0.91/245   \\
     &         & all  & bb + gauss$^1$ & 92$\pm$4     & 1.12$^2$      & 413$\pm$19 & $-56\pm$13 & 0.82/243   \\
\hline\noalign{\smallskip}
\rxa & EPIC-pn & all  & black-body     & 44.9$\pm$1.3 & 1.02$\pm$0.26 &    --      &    --      & 1.85/51   \\
     &         & all  & bb + gauss$^1$ & 45.0$\pm$2.6 & 2.02$\pm$0.71 & 329$\pm$24 & $-43\pm$3  & 0.92/49   \\
\noalign{\smallskip}
     &         & soft & black-body     & 44.0$\pm$1.6 & 1.02$\pm$0.32 &    --      &   --       &           \\
     &         & hard & black-body     & 45.7$\pm$1.6 & =             &    --      &   --       & 1.84/76   \\
     &         & soft & bb + gauss$^1$ & 44.5$\pm$1.3 & 2.09$\pm$0.42 & 329$^2$    & $-45\pm$10 &           \\
     &         & hard & bb + gauss$^1$ & 45.6$\pm$1.2 & =             & 329$^2$    & $-46\pm$8  & 1.17/74   \\
\noalign{\smallskip}\hline
\end{tabular}
\end{center}

$^1$Black-body model with Gaussian absorption line with fixed width of $\sigma$=70 eV. $^2$Parameter is fixed in the fit.
\label{tab-mresults}
\end{table*}

The four EPIC-pn spectra of \rxa\ were analyzed in a similar way as those of \rxc.
Due to the lower flux of this target we allowed only the black-body normalization to vary between the
spectra of the individual observations and fitted temperature and \nh\ as common parameters. The
results are presented in Table~\ref{tab-sresults}.
As for \rxc\ the four spectra are consistent with each other allowing to accumulate a total spectrum
from the merged data (Fig.~\ref{epic-spectra}). The fit results are listed in Table~\ref{tab-mresults}.
The temperature of 45 eV is the lowest value derived from any of the known XDINs.

\subsubsection{Absorption features ?}

The quality of the black-body fits to the
phase-averaged merged EPIC-pn spectra of \rxc\ and \rxa\ is in
both cases not very good. Recently, broad X-ray absorption features 
from the three XDINs \rbs\ \citep[=\rxd,][]{2003A&A...403L..19H}, 
\rxb\ \citep{2004A&A...419.1077H} and
\rxe\ \citep{2004ApJ.vanKerkwijk} were discovered in XMM-Newton
spectra. In the following we add a broad absorption line to the
black-body model. Because the statistical quality in the present spectra is
relatively low we fixed the $\sigma$ width of the line at 70 eV, a
value similar to those found from \rxb\ and \rxe. The best fit
parameters are summarized in Table~\ref{tab-mresults} (phases
`all') and the model fit is presented in Fig.~\ref{epic-spectra}.
Formally, the quality of the fit improves considerably in both cases when
including the absorption line. At the current stage of the EPIC-pn
spectral calibration it is not clear how much of this improvement
is caused by residual systematic calibration effects. In
particular, the residuals of the black-body fit to the spectrum of
\rxc\ are very similar in shape and strength to those of \rxf\
\citep{2004A&A...419.1077H}. On the other hand, for \rxa\ the
deviations are larger {\rm and} lead to a stronger absorption
feature (\eqw\ = $-$43 eV). At the derived energy of 329 eV \rxf\
does not show negative residuals which would lead to an artificial
absorption feature. The reduction in the (reduced) $\chi^2$ is also 
larger in the case of \rxa\ than in \rxf.

\subsubsection{Other models}

Following \citet{2003A&A...403L..19H} we also tried non-magnetic neutron star atmosphere
models \citep[e.g.][]{2002A&A...386.1001G,2002nsps.conf..263Z}. As in the case of \rbs\
iron and solar abundance atmospheres do not reproduce the smooth continuum observed from
\rxc\ and \rxa\ at energies above 0.5 keV where these models predict Fe absorption features.
The Hydrogen atmosphere model fits yield for both sources lower temperatures as the
black-body fits (35 eV for \rxc\ and 12 eV for \rxa) which leads to the well known problem
of predicting too high optical fluxes \citep{1996ApJ...472L..33P}. The quality of the
H atmosphere model fits range between those obtained from fits using a black-body with
and without line (reduced $\chi^2$ of 1.7 for 137 dof and 1.5 for 51 dof for
\rxc\ and \rxa, respectively).

Another model, first applied by \citet{2003A&A...399.1109B} to the
spectrum of \rxf\ and more recently by \citet{2004A&A...415L..31D} to
spectra of \rxb, consists of a black-body component attenuated by
photo-electric absorption and an additional ad hoc term
$\propto$ E$^{\beta}$. This model can successfully describe the
long-term spectral variations discovered in XMM-Newton RGS spectra
from \rxb\ with the value of $\beta$ increasing with time (over
nearly 2.5 years) from 1.4 to 2.7. The reported value for \rxf\ of
1.28$\pm$0.3 is at the lower end of values measured from \rxb. A
fit of this model to the EPIC-pn spectrum of \rxc\ results in an
increased temperature of 130 eV, an increased \nh\ of 1.9\hcm{20}
and $\beta$ = $-$1.76, very different to the values reported from
the high resolution spectra of \rxf\ and \rxb. Formally, this
model (with one more free degree of freedom) allows a better
representation of the continuum curvature (reduced $\chi^2$ =
1.50) but does not improve the residual pattern seen on smaller
energy scales in the pure black-body fit (Fig.~\ref{epic-spectra}
upper left). Hence, the best fit is obtained by including both,
power-law attenuation and a narrower absorption feature (reduced
$\chi^2$ of 1.0 for 134 dof). However, given the systematic
calibration uncertainties it is not clear if the broad-band
black-body model modification is justified in the case of \rxc.
For \rxa\ the extremely soft spectrum covers a very narrow energy
band which does not allow to constrain kT and $\beta$. Formally,
the best fit yields $\beta$ = -9.0 and kT = 550 eV. We therefore
do not follow this model further for this source.

\subsubsection{RGS spectra}

\begin{figure*}
\resizebox{8.7cm}{!}{\includegraphics[clip=]{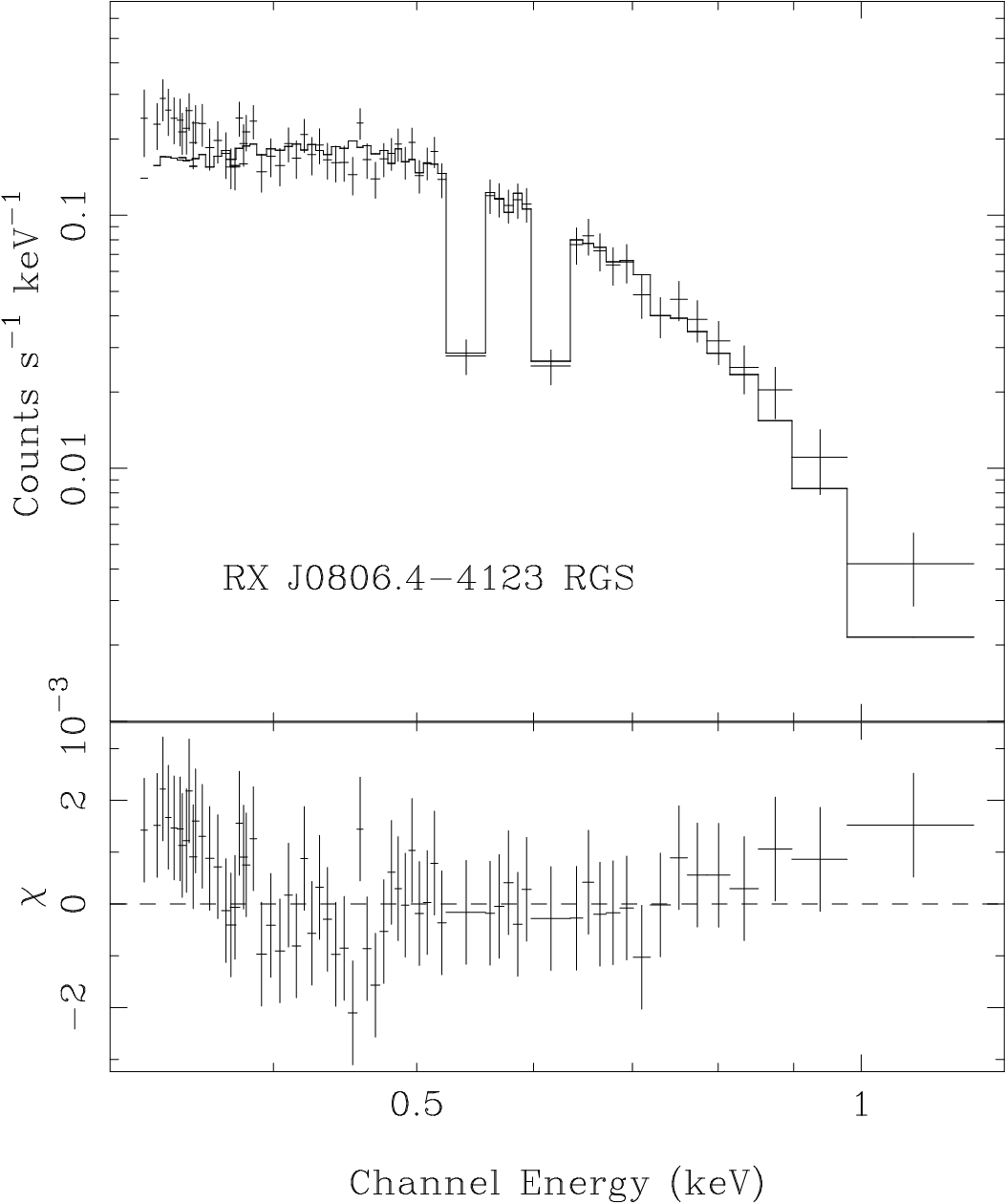}}
\hspace{2mm}
\resizebox{8.7cm}{!}{\includegraphics[clip=]{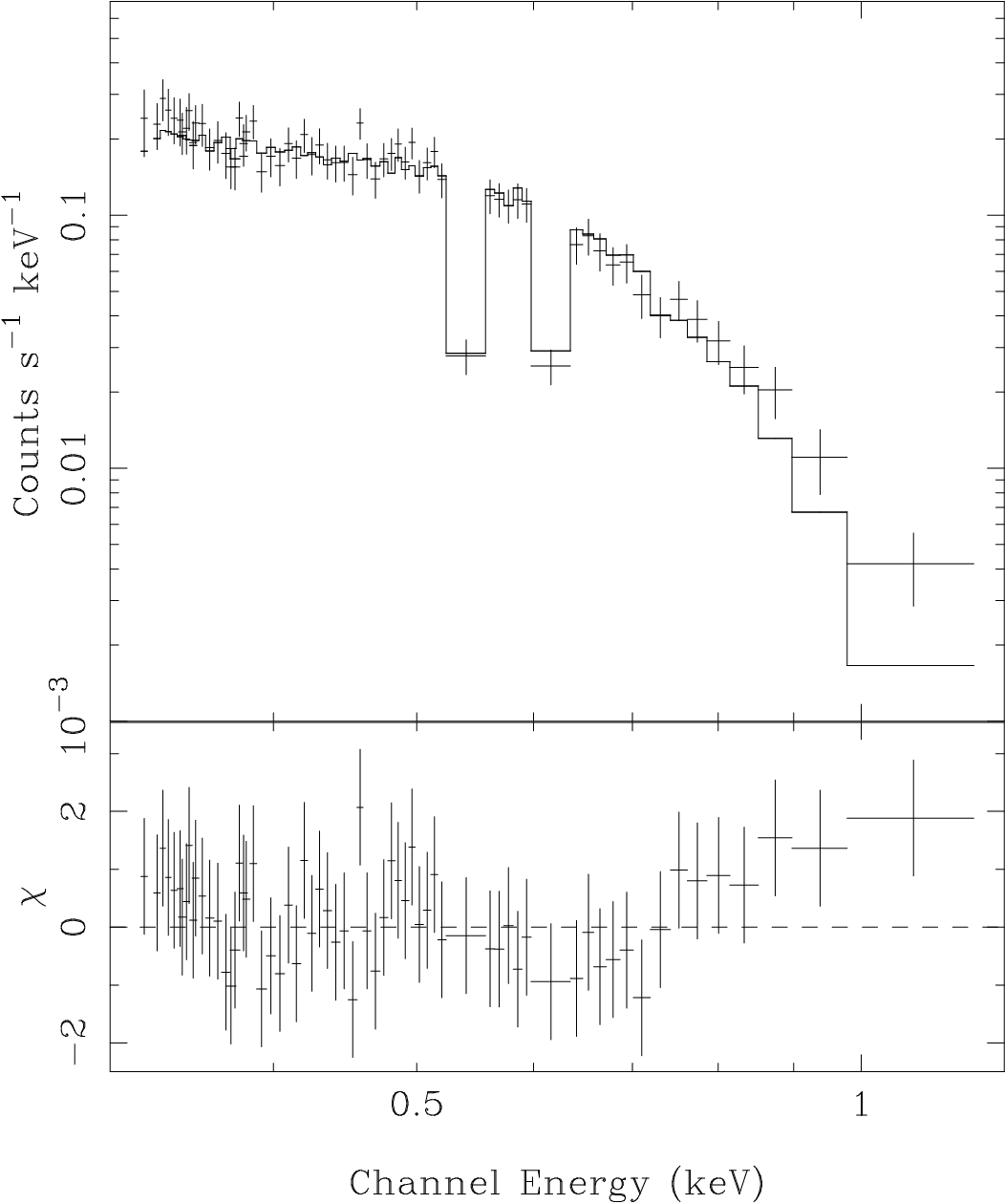}}

\caption{RGS spectrum obtained by combining RGS1 and RGS2 spectra from the two 
         observations of \rxc. As in Fig.~\ref{epic-spectra} the 
	 best fit black-body model with (right) and without (left) absorption line 
	 is plotted.}
\label{rgs-spectra}
\end{figure*}

RGS spectra of \rxc\ were produced using `rgsproc' of SAS 6.0. The four
spectra (RGS1 and RGS2 from the two observations) were simultaneously fit (only allowing
for individual normalization factors) using
the black-body model with and without including a broad absorption line.
Because of the lower sensitivity to photo-electric absorption in the RGS
spectra \nh\ was fixed to the values obtained from the EPIC-pn spectra.
The resulting best-fit parameters are presented in Table~\ref{tab-mresults} and the, 
for clearer representation combined, RGS spectrum is shown in Fig.~\ref{rgs-spectra}.

Although the black-body model yields an acceptable fit to the RGS spectra, deviations 
are visible at low energies. The residuals are improved when the absorption line 
is added to the model. Different values for the line energy and the line strength
are derived from RGS and EPIC-pn spectra, which most likely can be attributed to 
calibration problems.
 
\subsection{X-ray pulsations}

Due to the high time resolution of the EPIC-pn read-out mode only the data
from this instrument were used for our temporal analysis.
We extracted pattern 0-12 events from circular regions around the sources with
30\arcsec\ (\rxa) and 45\arcsec\ (\rxc) radius.
The event arrival times were checked for time jumps which sometimes occur in the pn data
and corrected to solar system barycenter using the `barycen' task.
For periodicity searches we followed HZ02
using the $Z^2_n$ method \citep{1983A&A...128..245B} 
and the Odds-ratio approach based on the Bayesian formalism
\citep{1996ApJ...473.1059G} 
for a precise determination of the spin period and its error 
\citep[see][ for details]{2004ApJ...606..444Z}.

\begin{figure*}
\resizebox{8.7cm}{!}{\includegraphics[angle=-90,clip=]{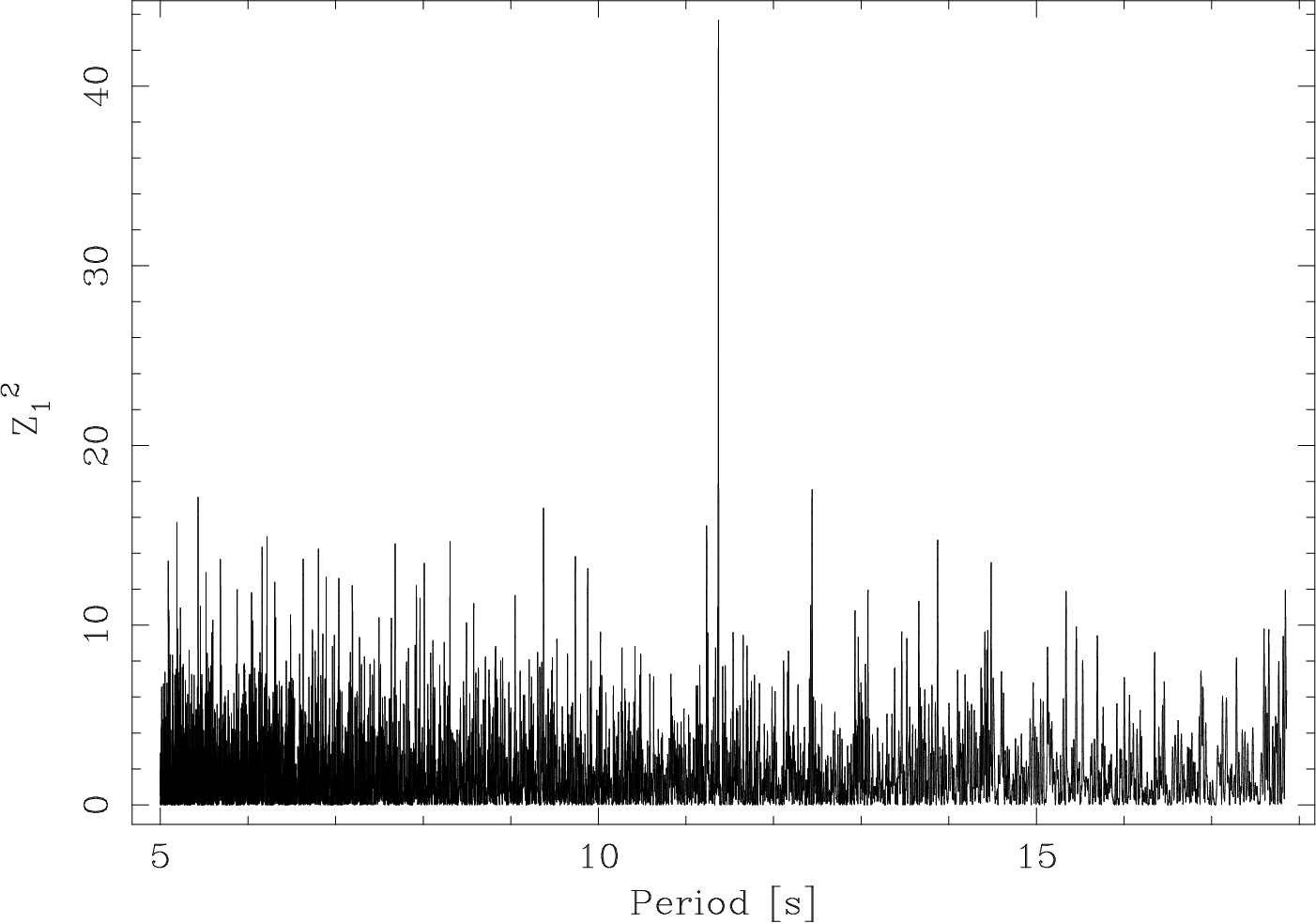}}
\hspace{2mm}
\resizebox{8.7cm}{!}{\includegraphics[angle=-90,clip=]{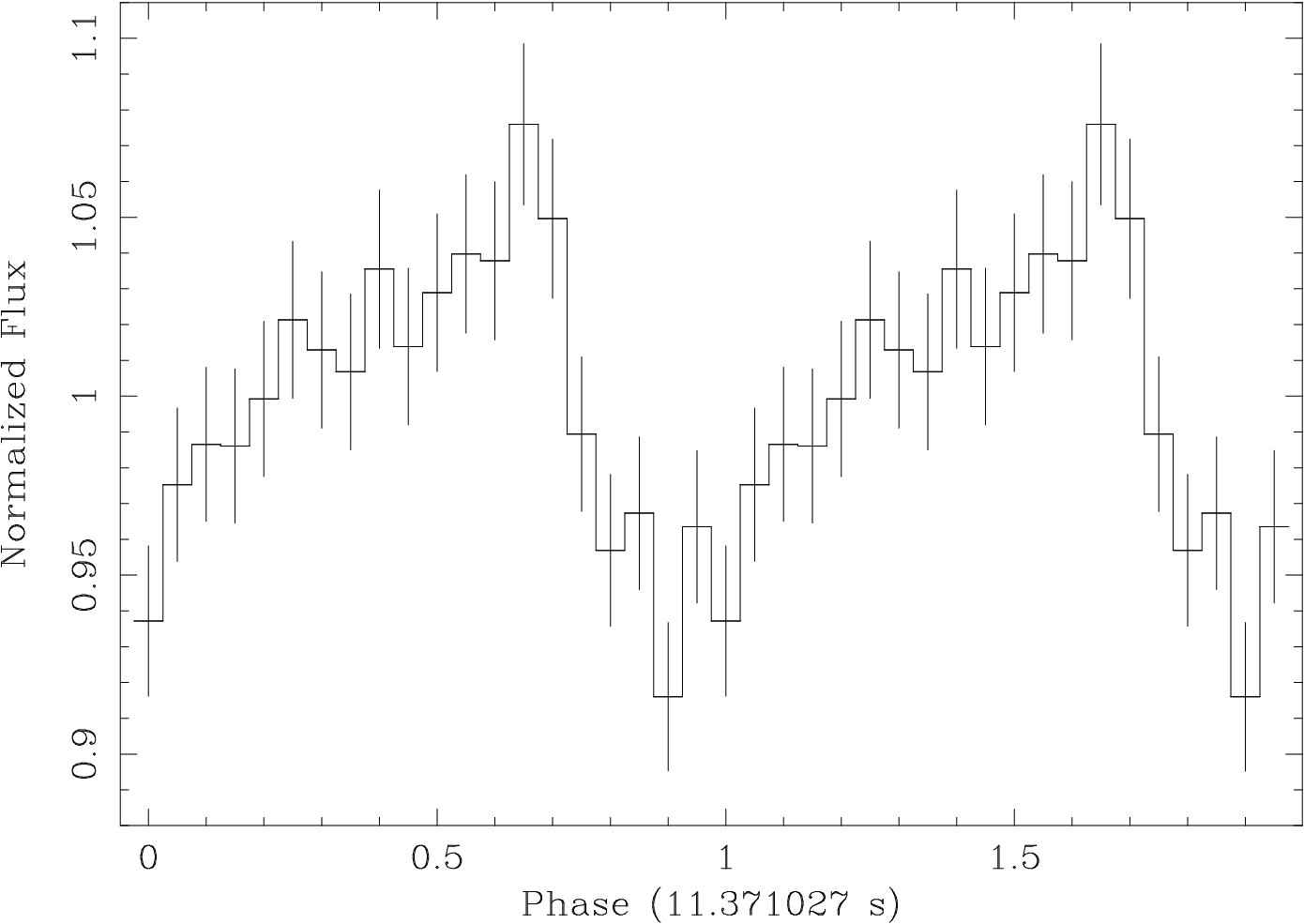}}
\caption{Left: Periodogram derived from a $Z^2_1$ test using the 0.12-1.2 keV EPIC-pn data of \rxc\ obtained in
satellite revolution 618 (April 2003). Right: Light curve folded at the best period, normalized to the average
count rate.}
\label{rxc-pulse}
\end{figure*}

For both targets neutron star spin periods were reported in the past with low significance:
11.371 s at a 3.5$\sigma$ level for \rxc\ from EPIC-pn data (HZ02) and evidence
for 22.7 s in ROSAT HRI data of \rxa\ \citep{1999A&A...351L..53H}. A search for pulsations around
11.371 s in the data of the second XMM-Newton observation of \rxc\ confirms the pulse period
and the $Z^2_1$ periodogram and the pulse profile are presented in Fig.~\ref{rxc-pulse}.
We summarize in Table~\ref{tab-tresults} the pulse periods with errors derived from the Odds-ratio method and the
$Z^2$ values for \rxc\ with the values for the first observation taken from HZ02 for comparison.

\begin{figure*}
\vspace{3mm}
\resizebox{8.7cm}{!}{\includegraphics[angle=-90,clip=]{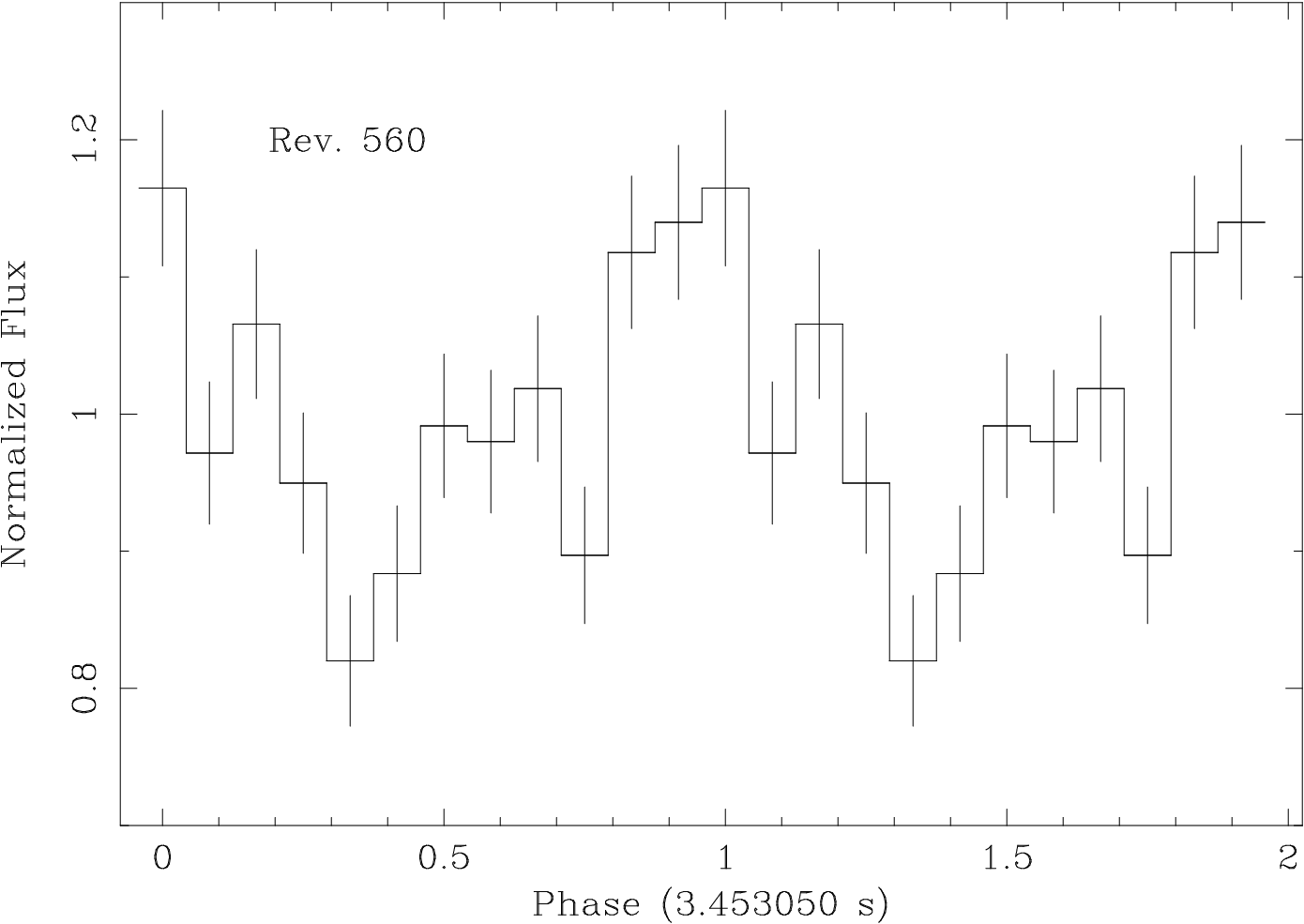}}
\hspace{2mm}
\resizebox{8.7cm}{!}{\includegraphics[angle=-90,clip=]{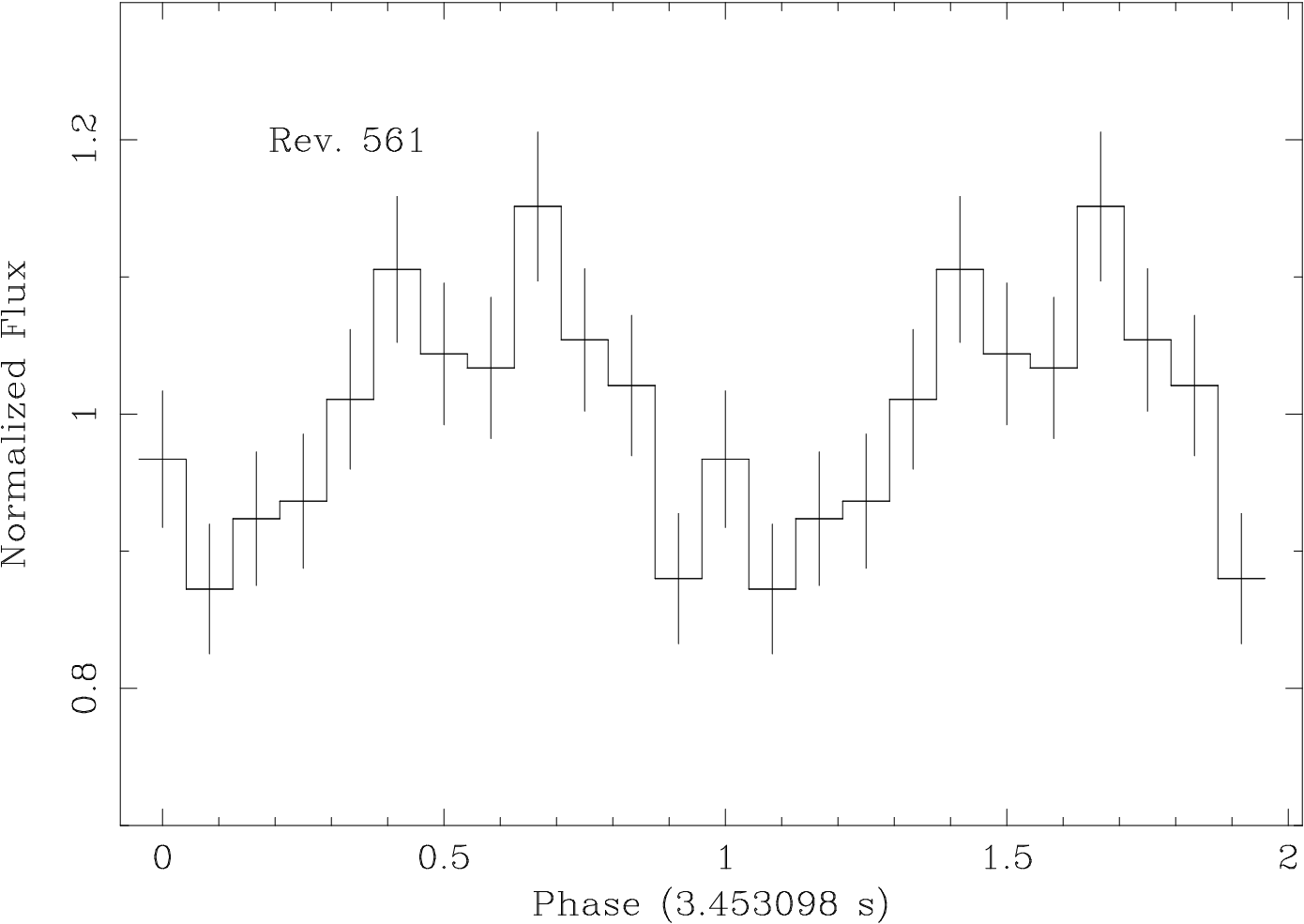}}

\vspace{3mm}
\resizebox{8.7cm}{!}{\includegraphics[angle=-90,clip=]{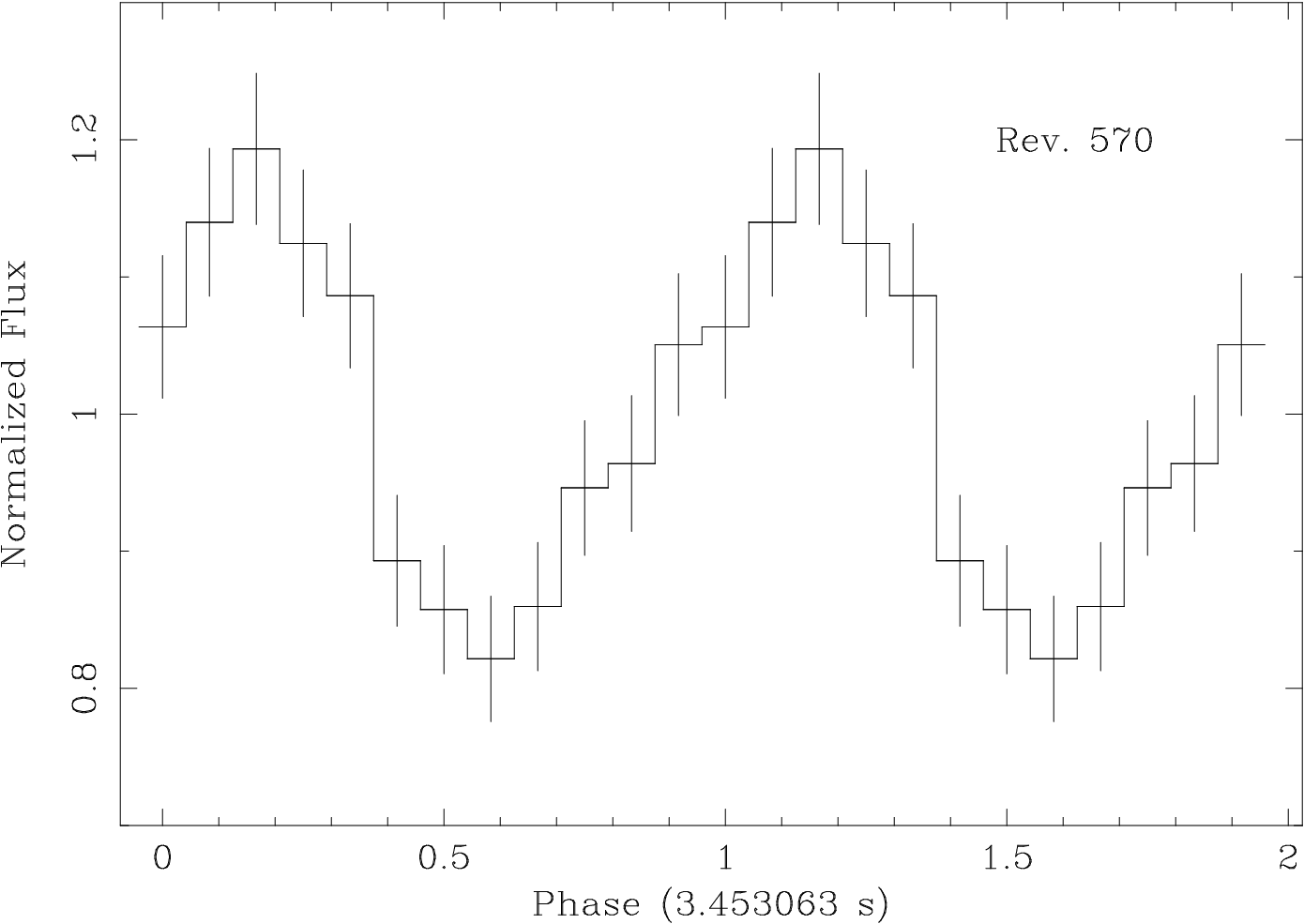}}
\hspace{2mm}
\resizebox{8.7cm}{!}{\includegraphics[angle=-90,clip=]{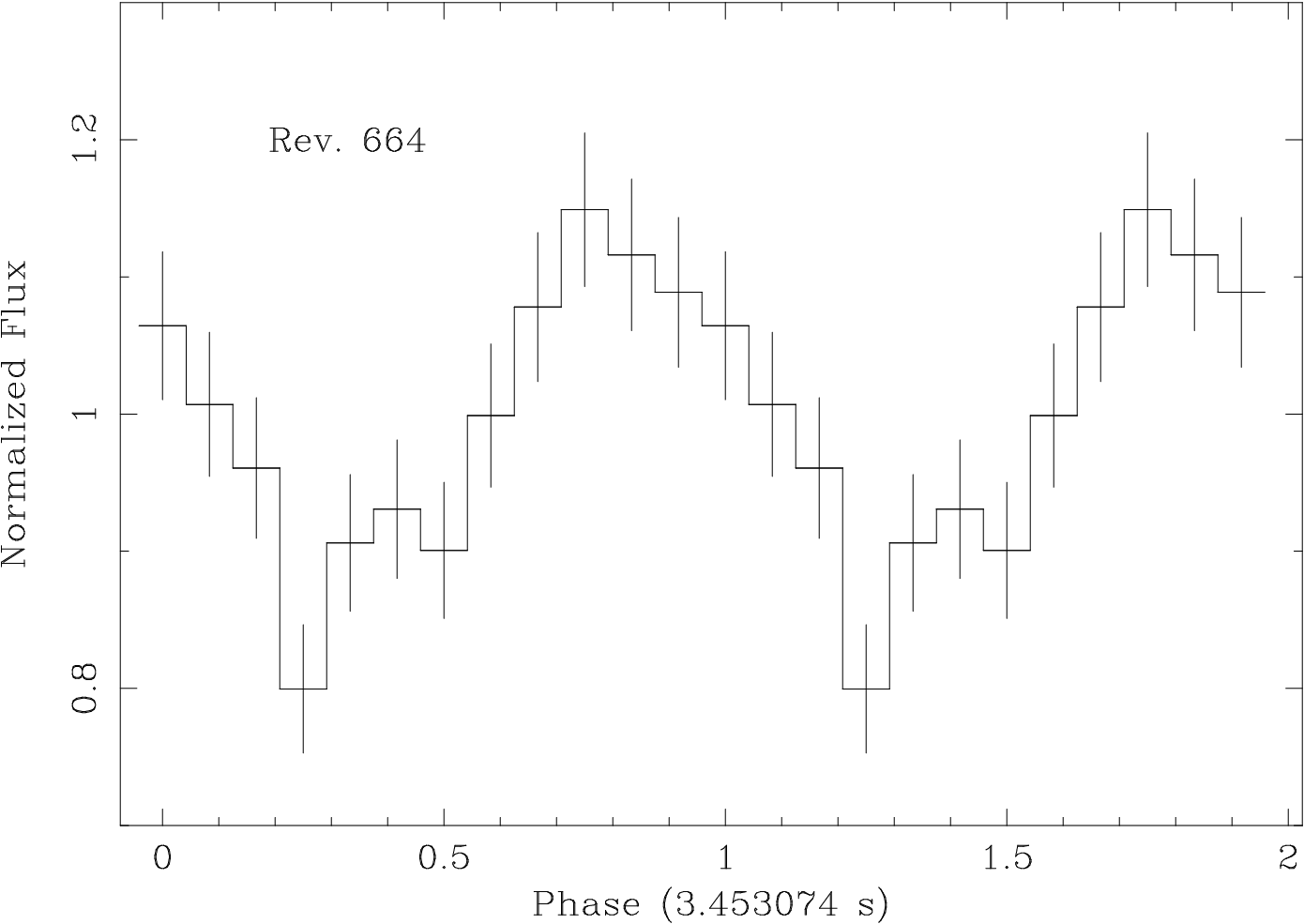}}
\caption{EPIC-pn 0.12-0.7 keV folded light curves of \rxa\ obtained from the four observations.
         The relative phases are arbitrary.}
\label{rxa-pulse}
\end{figure*}

None of the four observations of \rxa\ revealed a significant peak in the periodograms around the
suggested ROSAT value of 22.7 s. However, in all four observations a period of 3.453 s was detected.
As for \rxc, 
we performed $Z^2$ tests with contributions of higher harmonics. The results are listed in 
Table~\ref{tab-tresults} with the best pulse periods and errors. 
The pulse periods are consistent within the 1$\sigma$ errors.
The pulse profiles, folded at the best period and normalized to the mean count rates are shown in
Fig.~\ref{rxa-pulse}. Folding the light curves at the same average period yields no significantly 
different profiles. 
Although for all four observations the exposure, background level and net source intensity were
nearly identical, the maximum $Z^2_1$ value varied from one observation to the next.
However, the standard deviation for the $Z^2_1$ value expected from a sine-like signal is 
2\,$\sqrt{Z^2_1}+1$, indicating that the variations are not significant. 
To verify this we performed an additional
$\chi^2$ test following equation (1) of \citet{1998A&A...329..583Z} by comparing the observations
pair by pair. Since the relative phasing is not known we stepped through a grid of phase shifts
and used the minimum $\chi^2$ value to calculate the probability that the two light curves have different shapes.
The $\chi^2$ values and inferred probabilities are summarized in Table~\ref{tab-presults}. The highest
probability (98.5\% corresponding to 2.4$\sigma$), which was found from the comparison of the first and second
observation, is still consistent with statistical fluctuations. We conclude that there is no
significant change in the pulse profile of \rxa\ over 0.5 years covered by the observations.

\begin{table}
\caption[]{Results of the temporal analysis of EPIC-pn data.}
\begin{center}
\begin{tabular}{lccccc}
\hline\hline\noalign{\smallskip}
\multicolumn{1}{l}{Obs.} &
\multicolumn{1}{c}{Period} &
\multicolumn{1}{c}{$Z^2_1$} &
\multicolumn{1}{c}{$Z^2_2$} &
\multicolumn{1}{c}{$Z^2_3$} &
\multicolumn{1}{c}{$Z^2_4$} \\

\multicolumn{1}{l}{} &
\multicolumn{1}{c}{[s]} &
\multicolumn{1}{c}{} &
\multicolumn{1}{c}{} &
\multicolumn{1}{c}{} &
\multicolumn{1}{c}{} \\

\noalign{\smallskip}\hline\noalign{\smallskip}
\multicolumn{6}{l}{\rxc} \\
168-pn & 11.37139$\pm$3.0\expo{-4} & 33.9 & 42.9 & 43.0 & 43.2 \\
618-pn & 11.37103$\pm$4.0\expo{-5} & 43.6 & 57.2 & 60.4 & 60.9 \\
\hline\noalign{\smallskip}
\multicolumn{6}{l}{\rxa} \\
560-pn & 3.453050$\pm$6.0\expo{-5} & 30.0 & 36.9 & 48.5 & 51.7 \\
561-pn & 3.453098$\pm$7.2\expo{-5} & 27.2 & 30.1 & 32.1 & 33.9 \\
570-pn & 3.453063$\pm$3.5\expo{-5} & 67.0 & 69.0 & 70.3 & 72.1 \\
664-pn & 3.453074$\pm$8.4\expo{-5} & 36.4 & 36.5 & 39.0 & 41.5 \\
\noalign{\smallskip}\hline
\end{tabular}
\end{center}

Errors on the period are given for a 68\% confidence level.
\label{tab-tresults}
\end{table}

\begin{figure*}
\resizebox{8.7cm}{!}{\includegraphics[clip=]{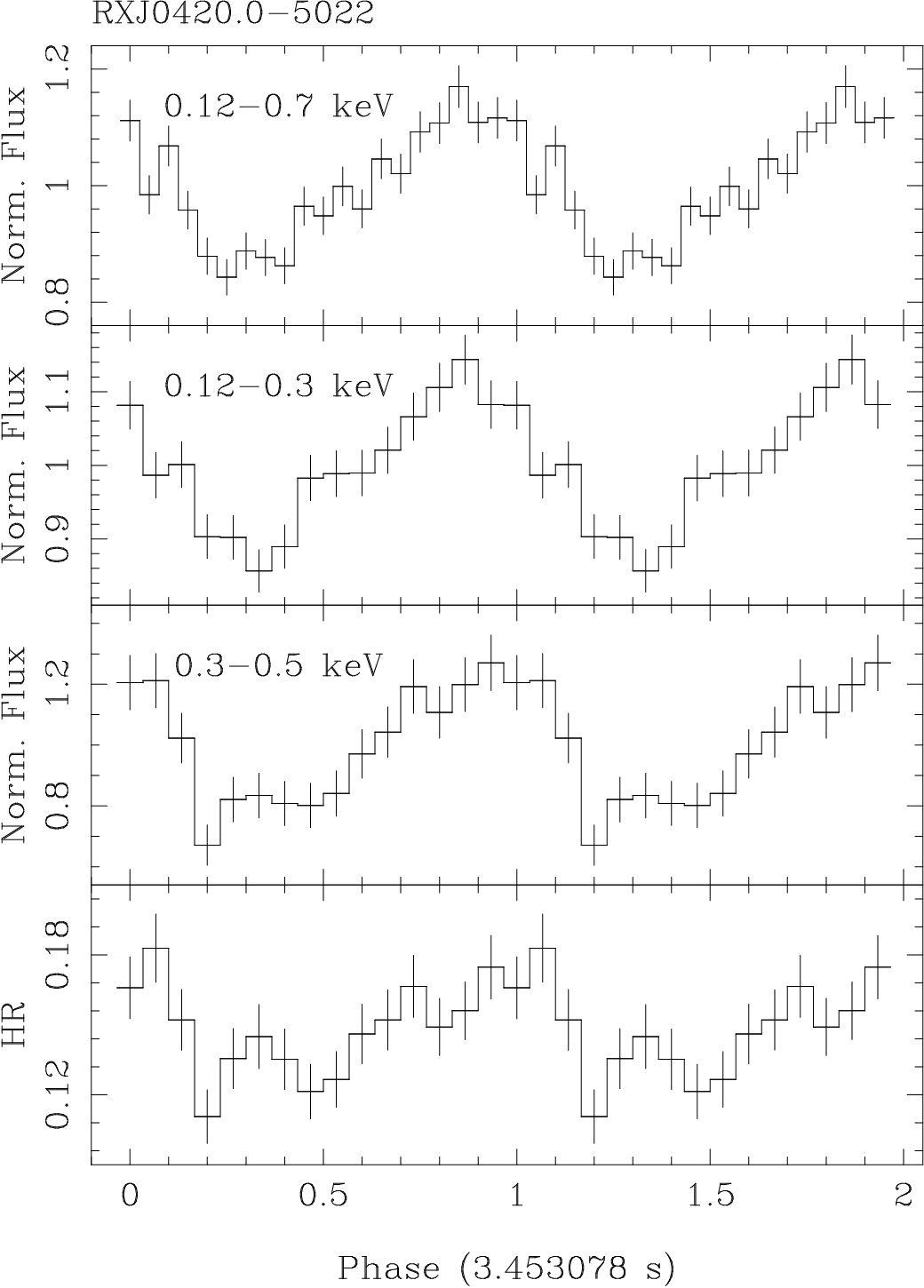}}
\hspace{2mm}
\resizebox{8.7cm}{!}{\includegraphics[clip=]{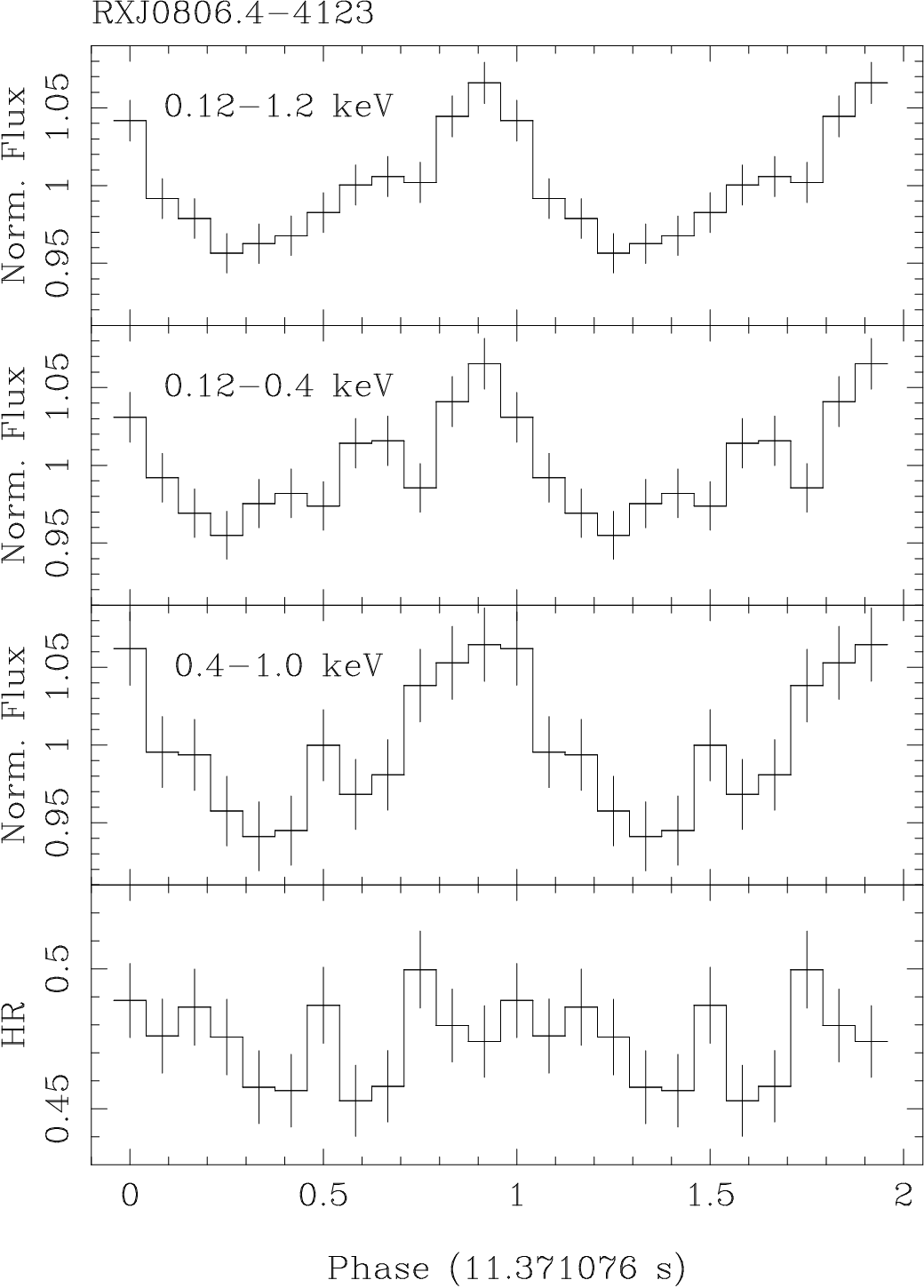}}
\caption{Folded light curves in the total energy band covered by the spectrum and in two
         sub-bands together with the hardness ratio, the ratio of the
         count rates in the hard to the soft band.}
\label{epic-hardness}
\end{figure*}

The time baseline between the observations is not sufficiently long to detect significant period changes
for both objects. Fitting a linear period change with time we derive spin period change rates \pdot\ of
(-4.6$\pm$6.4)\expo{-12} s s$^{-1}$ and (0.5$\pm$8.6)\expo{-12} s s$^{-1}$ for \rxc\ and \rxa, respectively.
The error on the spin period of \rxa\ accumulates to nearly one period over the 2 day gap between
the first two observations. This is too large to be able to phase-connect the observations as it was originally
one aim of the proposal where the gaps between the observations were optimized for a spin period of 22.7 s.

\begin{table}
\caption[]{Comparison of pulse profiles for \rxa.}
\begin{center}
\begin{tabular}{lrrrr|c}
\hline\hline\noalign{\smallskip}
\multicolumn{1}{l}{Revolution} &
\multicolumn{1}{c}{560} &
\multicolumn{1}{c}{561} &
\multicolumn{1}{c}{570} &
\multicolumn{1}{c}{664} &
\multicolumn{1}{|c}{PF [\%]} \\

\noalign{\smallskip}\hline\noalign{\smallskip}
560  & $-$    & 21.97  & 17.57  & 10.99 & 10.2$\pm3.5$ \\
561  & 98.5\% & $-$    &  3.68  & 14.36 & 10.0$\pm3.5$ \\
570  & 93.7\% & 3.9\%  &  $-$   &  8.61 & 17.0$\pm3.5$ \\
664  & 64.2\% & 84.3\% & 43.1\% &  $-$  & 13.8$\pm3.5$ \\
\noalign{\smallskip}\hline
\end{tabular}
\end{center}

The upper right part of the table gives the $\chi^2$ values derived from the comparison
of the two pulse profiles for 10 degrees of freedom, dof (12 phase bins with two parameters for
the normalization). The corresponding probability that the two light curves of a pair 
have different shapes is given in the lower left part.
\label{tab-presults}
\end{table}

\subsection{Pulse phase resolved X-ray spectra}

For \rxa\ no significant changes of pulse period and pulse shape
were detected. Therefore, we merged the data of the four
observations and performed a $Z^2_1$ test to determine a medium
period assuming \pdot\ = 0. Light curves folded with the derived
period of 3.453078 s were then produced for two different energy
bands and a hardness ratio computed as ratio of the two light
curves (hard band / soft band). These are shown in
Fig.~\ref{epic-hardness} (left). 
For the 0.12-0.7 keV light curve we derive (by fitting 
a sine wave to the pulse profile) a pulsed fraction of 12.7$\pm$1.8\%. 
The pulsed fractions for the individual observations are reported in 
the last column of Table~\ref{tab-presults}.
Variations in
the hardness ratio (a $\chi^2$ test for a constant hardness
results in $\chi^2 = 34.6$ for 14 dof, corresponding to a
3.2$\sigma$ result) are visible with the hardness ratio maximum
delayed with respect to the intensity maximum, very similar to the
behaviour of \rxb\ \citep{2001A&A...365L.302C,2004A&A...419.1077H}.
The EPIC observations of \rxa\ cover only 0.5 years and do
not provide sufficient statistics to detect changes in the
relative phasing between flux and hardness ratio, as reported for
\rxb\ by \citet{2004A&A...415L..31D}.

\begin{figure}
\resizebox{8.8cm}{!}{\includegraphics[angle=-90,clip=]{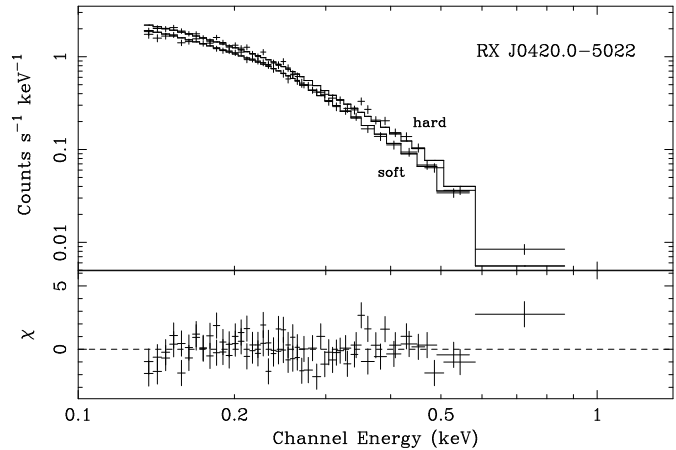}}
\caption{Phase resolved EPIC-pn spectra of \rxa\ fitted with a black-body plus Gaussian absorption line model.}
\label{epic-phase}
\end{figure}

To further investigate the spectral variations with pulse phase we accumulated spectra of \rxa\ from
phases around hardness maximum and minimum (phases 0.6-1.1 and 0.1-0.6 as used in Fig.~\ref{epic-hardness}).
The two spectra were again fitted with absorbed black-body model with and without absorption line.
The results are summarized in Table~\ref{tab-mresults}. As in the case of the phase-averaged spectrum
adding the absorption line strongly improves the fit. There is an indication for an increased temperature
during the phases of higher hardness ratio, but the errors are large and more statistics is required to
prove the significance of this result. In Fig.~\ref{epic-phase} the spectra are presented with the best fit
model including an absorption line.

In the corresponding analysis of the merged data of \rxc\ the light curves, folded on the derived period
of 11.371076 s are shown in Fig.~\ref{epic-hardness} (right). \rxc\ exhibits the smallest pulsed fraction
of the known XDIN pulsars of about 6 \% which renders the detection of significant hardness ratio
variations difficult. The fit of constant hardness ratio results in $\chi^2 = 14.5$ for 11 dof which
corresponds to a probability of 79.3\% ($\sim1.3\sigma$). However, it should be noted that by assuming a
constant pulse period one may smear out the pulse profiles somewhat if the pulsar actually exhibits a
small period change over the 2.5 years from one observation to the next. Because of the marginal
variations in hardness ratio we do not perform a phase resolved spectral analysis for \rxc.

\section{Optical observations}

Deep optical CCD images of the field containing \rxa\ were
obtained under photometric conditions on November 22, 2000 with FORS1 at UT1
of the ESO VLT. Three CCD images with individual exposure times of 1200 s were
 taken through a Bessel B filter and two images with 600 s exposure time
each through a Bessel V filter. Two bright stars in the field have been masked
by movable slitlets to avoid saturation of the detector and charge spilling.
The mean seeing was 0.9\arcsec\ FWHM.

The images were reduced and averaged using bias images and sky flats
obtained during the same observing run and with the same instrument and
detector setups. Cosmic ray events were removed by applying the FILTER/COSMIC
command of the ESO MIDAS software package to the averaged B- and V-filter
images. An astrometrical calibration was derived from the positions of
21 field stars from the USNO-B catalog in the vicinity of \rxa.

\begin{figure}
\resizebox{8.8cm}{!}{\includegraphics[clip=]{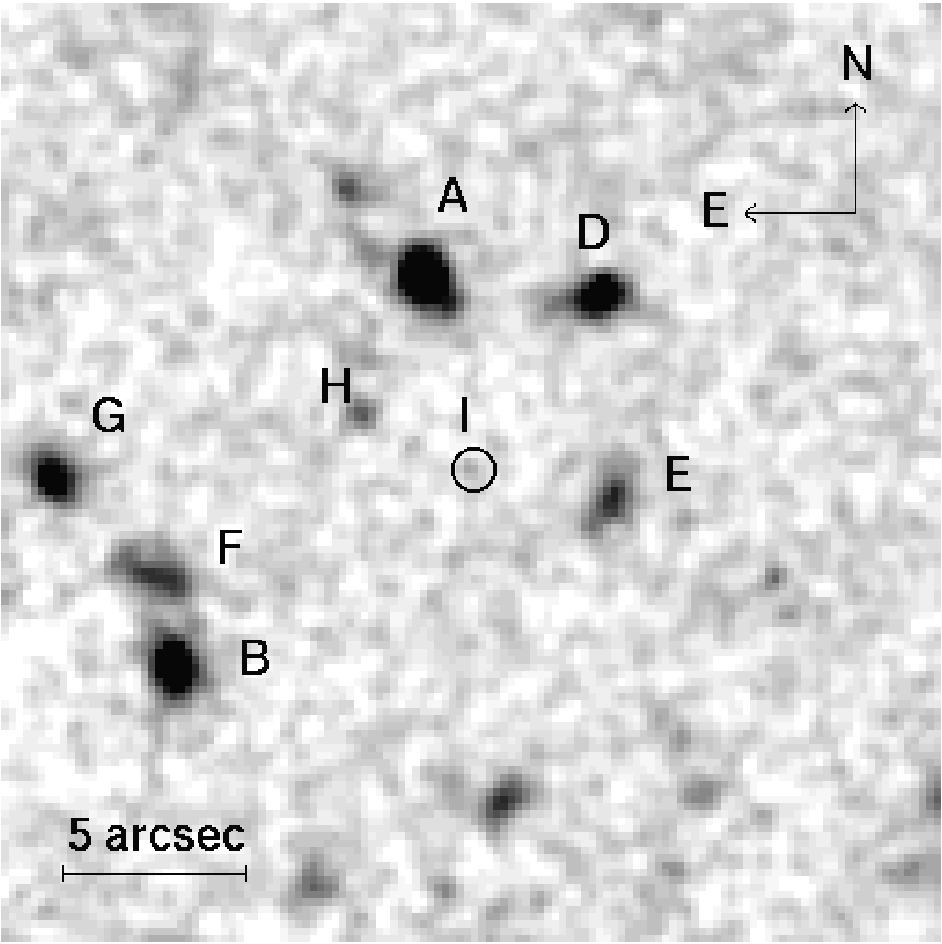}}
\caption{Averaged and smoothed B-filter image obtained with FORS1 at the ESO VLT
of the field of \rxa\ with a total exposure of 1\,h.  The circle with
0.6\arcsec\ radius marks the Chandra position of the X-ray source and includes
a possible optical counterpart with B = 26.57$\pm$0.30 mag.}
\label{fig-vlt}
\end{figure}

Within the error circle of the Chandra X-ray position of \rxa\
a single object is marginally detected on the B-filter image (Fig.~\ref{fig-vlt})
with B = 26.57$\pm$0.30 mag.
No object was found at this position on the V-filter image down to a detection
limit V $>$ 25.5.
Fluxes for objects A, D, E, and H from \citet{1999A&A...351L..53H}, the new
candidate star I, and the local sky background were determined using
aperture photometry.
For the photometric calibration nightly zero points and extinction coefficients
were used from the FORS1 quality control pages at ESO.
The resulting B and V magnitudes are given in Table~\ref{tab-vlt}. The new B band
photometry of stars A, D, E, and H is fully consistent with that given in
\citet{1999A&A...351L..53H}.
The detection and photometry of star I is adversely affected by the locally
enhanced sky background due to the stray light halo of an 11 mag star at
$\sim 45\arcsec$ distance. The error uncertainty given for the B magnitude
is the statistical error calculated from the photon statistics of the object
and background fluxes. It does not include systematic uncertainties in the
determination of the background flux which could be of similar order.

\begin{table}
\caption{Optical photometry of candidate stars in the vicinity of \rxa.}
\begin{center}
\begin{tabular}{ll@{$\pm$}rl@{$\pm$}r}
\hline\hline\noalign{\smallskip}
Object & \multicolumn{2}{c}{B [mag]} & \multicolumn{2}{c}{V [mag]} \\
\noalign{\smallskip}\hline\noalign{\smallskip}
A & 24.40 & 0.04 & 23.46 & 0.03 \\
D & 24.88 & 0.06 & 24.38 & 0.07 \\
E & 25.33 & 0.10 & 24.74 & 0.10 \\
H & 26.11 & 0.20 & 24.95 & 0.13 \\
I & 26.57 & 0.30 & \multicolumn{2}{c}{$>$25.5} \\
\noalign{\smallskip}\hline
\end{tabular}
\end{center}
\label{tab-vlt}
\end{table}

\section{Discussion}

XMM-Newton observations of the two isolated neutron stars \rxa\
and \rxc\ establish that both are X-ray pulsars with spin periods
of 3.453 s and 11.371 s, respectively. \rxa\ exhibits a relatively
strong sinusoidal intensity modulation of $\sim$12\% similar 
to that observed from \rxb\
\citep{1997A&A...326..662H,2001A&A...365L.298P}. Also, like for
\rxb, the hardness ratio shows variations with pulse phase with
maximum (hardest spectrum) shifted with respect to the intensity
maximum \citep{2001A&A...365L.302C,2004A&A...419.1077H,2004A&A...415L..31D}.

The light curve of RX J0806.4-4123 (Fig.~\ref{rxc-pulse}) shows a possible
deviation from a purely sinusoidal shape, with a relatively
smooth rise of the pulse followed by a steeper decline. The same
behaviour may also be present (although to a lower level of
confidence) in the light curve of RX J0420.0-5022 
(Fig.~\ref{rxa-pulse}), 
and is different from what
is exhibited by \rxb. The pulse profile of the latter
source appears, in fact, quite symmetric with respect to the
pulse maximum/minimum \citep{2004A&A...419.1077H}.  An asymmetric
(skewed) pulse profile hints towards a surface temperature
distribution more complex than the (symmetric) one implied by a
simple dipolar magnetic field \citep{1995ApJ...442..273P,1996ApJ...473.1067P}. 
If what we are observing is the thermal emission from the entire
star surface, the simplest explanation would be to invoke the
presence of higher order multipolar components in the magnetic
field of these objects.

The pulse phase averaged X-ray spectra of both pulsars
investigated here are not very well fit by an absorbed Planckian
model. The derived black-body temperature of 96 eV for \rxc\ is
among the highest observed from this group of ROSAT discovered
isolated neutron stars. The temperature kT of 45 eV inferred for
\rxa, which could only roughly be estimated from the ROSAT PSPC
spectrum \citep{1999A&A...351L..53H}, on the other hand, is the
lowest value observed to date \citet{haberl2004COSPAR}. Adding a
Gaussian-shaped absorption line to the continuum model 
improves the spectral fits for both objects,
suggesting that there may be broad absorption features present in
the X-ray spectra similar to \rbs\ \citep{2003A&A...403L..19H},
\rxe\ \citep{2004ApJ.vanKerkwijk} and \rxb\
\citep{2004A&A...419.1077H}. 
If interpreted as proton cyclotron absorption
lines \citep[as supported by the agreement between 
the magnetic field strengths independently inferred from spin-down
measurement and cyclotron energy in the case
of \rxb, see][]{2004MNRAS.Cropper} the
inferred energies of $\sim$330 eV and $\sim$460 eV would indicate 
magnetic field strengths of $\sim$6.6\expo{13} G and 
$\sim$9.2\expo{13} G for \rxa\ and \rxc, respectively.
However, because the residuals seen in the black-body fit to 
the EPIC spectra of \rxc\  are similar in shape to those seen from \rxf\
\citep{2004A&A...419.1077H} and the RGS spectra yield somewhat
different line parameters it is not clear to which extent they
are influenced by systematic calibration effects. 

It is remarkable that probably up to five out of the six X-ray dim
isolated neutron stars detected in the ROSAT all-sky survey, show
broad absorption lines with central energies in the range of
$\sim$200 eV \citep[only an upper limit of $\sim$300 eV was found
for \rbs,][]{2003A&A...403L..19H} to 450 eV
\citep[\rxe,][]{2004ApJ.vanKerkwijk}. An interpretation in terms
of fundamental lines of a proton cyclotron emission implies
surface magnetic fields in the relatively narrow range of B = 3-7
10$^{13}$ (1+$z$) G, with $z\simeq$\,0.3 the redshift parameter. 
These estimates are comparable with the QED critical value
B$_{\rm QED}\simeq$\,4.4\expo{13} G, that is expected to quench pulsar 
activity. The upper B value could simply reflect the lower sensitivity 
to line detection in the Wien part of the spectrum.
Were these stars active pulsars, the spin periods of the
order of 10 s observed in some sources imply that any radio pencil
beam should be very narrow \citep{1990MNRAS.245..514B} 
and the chances then high that it does not sweep over the Earth.  
Magnetic field strengths below the critical value
would still be compatible with the presence of cyclotron lines
if some of the detected features are higher order harmonics.
Unfortunately, little is known about
harmonics to fundamental line ratios. All opacities given so far
for proton cyclotron lines have been computed neglecting the
contribution of higher harmonics and the absence of such lines in
model atmospheres does not mean that they should not be
observable. In the illustrative but still
unique case of 1E\,1207.4-5209 \citep{2003Natur.423..725B,2004A&A...418..625D},
the line equivalent widths
of the various components (65 - 100 eV) do not vary much with central energy (690 to 2800 eV).
Harmonic lines with similar equivalent widths could be missed in the very soft spectrum of \rxa,
or maybe in the case of \rxe, with the highest line energy, but would have been detected
in \rbs\ and \rxb.

In the magnetic field and temperature conditions prevailing in XDINs, hydrogen atmospheres
could be partly ionized as a result of the much increased ionization potential at large B
\citep[see, e.g., ][ and references therein]{2002nsps.conf..263Z}.
Bound-free and bound-bound transitions occur at energies which can be made comparable to those of
the lines observed in XDINs. Since atomic line energies vary more slowly with magnetic
field than cyclotron lines do, a larger range of surface magnetic fields could in principle be
allowed.  The main constraint of this interpretation is that the proton cyclotron line should be
located outside of the observable energy range, i.e., at E $\leq$ 100 eV, implying surface magnetic
fields below 2$\cdot$10$^{13}$G.  Model atmospheres by \citet{2003ApJ...599.1293H} 
show that at B = 10$^{13}$G, the s0-s2 transition at E $\sim$ 180 eV and the H ionization edge at E $\sim$ 250 eV
could both contribute to some of the observed lines. At higher energies, heavier elements could
play a role \citep{2002ApJ...578L.133H}. 
On the other hand, at such strong magnetic fields ($>$3\expo{13} G) and rather low surface temperatures
($<$\oexpo{6} K) the neutron star surface may be in a solid or condensed
state. At present, only a first step has been undertaken in
studying radiative properties of such surfaces \citep{2004AdSpR..33..531Z,2004ApJ...603..265T}.

The precise X-ray position of \rxa\ obtained with Chandra allowed us to identify a possible optical counterpart
on a deep VLT B band image. The shorter V band exposure is not sensitive enough to provide colour information
which could be used to support the identification. However, the extrapolation of the X-ray spectrum into the
optical results in a factor of 12.3 (or smaller if our identification is wrong) lower flux than actually
is observed in the B band. Similar optical `excess'
in the spectral energy distribution is seen from all optical counterparts (and candidates) identified so far
(factor 14 for \rxe\ \citep{2003ApJ...588L..33K}, $<$4.9 for \rbs\ \citep{2002ApJ...579L..29K}, 2.4 with an
additional component increasing with wavelength for \rxb\ \citep{2003ApJ...590.1008K} and $\sim$7 for \rxf\
\citep[][based on the spectral parameters derived from the LETGS X-ray spectrum]{2003A&A...399.1109B}.
Understanding the overall spectral energy distribution of XDINs from radio to X-ray wavelength will be crucial
to derive a comprehensive model of the emission properties if this group of neutron stars.

\begin{acknowledgements}
The XMM-Newton project is supported by the Bundesministerium f\"ur Bildung und
For\-schung / Deutsches Zentrum f\"ur Luft- und Raumfahrt (BMBF / DLR), the
Max-Planck-Gesellschaft and the Heidenhain-Stif\-tung.
\end{acknowledgements}

\bibliographystyle{apj}
\bibliography{ins,general,myrefereed,myunrefereed,mytechnical}

\end{document}